\documentclass[aps,pre,eqsecnum]{revtex4}
\usepackage{hyperref}

\usepackage{amsmath}
\usepackage{amssymb}
\usepackage{epsfig}
\usepackage{natbib}
\newcommand*{\be}{\begin{equation}}
\newcommand*{\ee}{\end{equation}}
\begin{document}
\bibliographystyle{revtex}
\title{Dynamics of subpicosecond dispersion-managed soliton in a
 fibre:\break A perturbative analysis}
\author{E.V. Doktorov}
\email{doktorov@dragon.bas-net.by} \affiliation{B.I. Stepanov
Institute of Physics, 68 F. Skaryna Ave., 220072 Minsk, Belarus}

 \begin{abstract}
 A model is studied which describes a propagation of a
 subpicosecond optical pulse in dispersion-managed  fibre links.
 In the limit of weak chromatic dispersion management, the model
 equation is reduced to a perturbed modified NLS equation having a
 nonlinearity dispersion term. By means of the Riemann--Hilbert
 problem, a perturbation theory for the soliton of the
 modified NLS equation is developed. It is shown in the adiabatic
 approximation that there exists a unique possibility to suppress
 the perturbation-induced shift of the soliton centre at the cost
 of proper matching of the soliton width and
 nonlinearity dispersion parameter. In the next-order
 approximation, the spectral density of the radiation power
 emitted by a soliton is calculated.
 \end{abstract}
 \maketitle

 \section{Introduction}

 The dispersion management for short pulse propagation in optical
 fibres is the key current technology for ultrafast high-bit-rate
 communication lines
 \cite{Smith,GT96,GST,Kumar,Lak-1,Rev,Med,Sample,Malomed}. The
 basic idea behind the dispersion management borrowed from linear
 systems \cite{Kogel} consists in a compensation of the chromatic
 dispersion by means of periodically incorporating additional
 fibre sections with opposite sign of chromatic dispersion.
 Optical pulses in a fibre link with incorporated sections, the
 so-called dispersion-managed solitons (DM-solitons), exhibit a number
 of true-soliton-like properties, including elastic scattering and
 high stability. Such a behaviour is unexpected because the
 underlying equation, the nonlinear Schr\"odinger (NLS) equation
 with variable chromatic dispersion coefficient, is by no means
 integrable. A recent example demonstrating a great success of the
 DM technology is a development of the commercial DWDM
 communication line in Australia with 160 frequency channels and
 1.6 Tbit/s total data transmission rate over 3875 km.

 In treating analytically the DM NLS solitons, asymptotic methods
 have been proved to be especially effective \cite{GT96,GST,AB}.
 Using the multiscale asymptotic expansion, Gabitov and Turitsyn
 \cite{GT96} decomposed the pulse dynamics into rapidly varying
 phase and slowly evolving amplitude and derived an
 integro-differential equation (the Gabitov--Turitsyn equation) for
 the pulse amplitude. It was shown \cite{Zakh,Lvov,LP}
 that in the limit of weak dispersion management this equation is
 reduced to the perturbed NLS equation.

 Up to now, DM solitons in fibres are considered within the
 NLS-based systems, i.e., for picosecond optical pulses. The
 demand for an increased bit rate of a communication line can be
 implemented by using pulses of subpicosecond duration ($\sim$100
 fs). For ultrashort pulses, the NLS equation becomes inadequate
 because more subtle effects, the main ones being the nonlinearity
 dispersion, the Raman self-frequency shift and linear third-order
 chromatic dispersion, come into play. It is important
 that, even for small strengths of these new effects, some of them cannot
 in general be considered as perturbations of the NLS equation. It
 is precisely such a situation that arises when attempting to account
 for the nonlinearity dispersion. On the one hand, it is well known
 \cite{Kiv} that the NLS soliton being subjected to the action of
 the nonlinearity dispersion, exhibits the so-called
 self-steepening. On the
 other hand, the NLS equation with the additional nonlinearity
 dispersion term, the modified NLS (MNLS) equation, is still
 integrable and possesses its own solitons that propagate without
 distortion. The reason for such a difference lies in the fact
 that the true MNLS soliton is \emph{non-perturbative} with
 respect to the nonlinearity dispersion parameter and cannot be
 obtained from the NLS equation with the nonlinearity dispersion
 term as a perturbation.

 In the present paper we give a self-contained exposition of the
 MNLS soliton dynamics in fibre links with variable chromatic
 dispersion. In section 2 we derive the Gabitov-Turitsyn-like
 integro-differential equation for the soliton amplitude and
 reduce it to a perturbed MNLS equation in the limit of weak
 chromatic dispersion management. In section 3 we construct the
 soliton solution of the MNLS equation using the Riemann--Hilbert
 (RH) problem. The MNLS soliton solution has been previously
 derived by different methods \cite{Rangwala,D-JMP,Zabol}. The
 representation for the MNLS soliton given here is the most simple
 and transparent. Section 4 is devoted to a development, within
 the RH problem approach, of a perturbation theory for the MNLS
 soliton. Note that the applicability of the RH problem for
 treating soliton perturbations has been initiated in papers
 \cite{Yura,Zhenya}. We considerably simplify the formalism and
 final expressions, as compared with the previous results
 \cite{D-V-PD,Yang,Lashkin} on the perturbed MNLS soliton. In
 section 5 we work out in detail the general results of the
 perturbation theory for the simplest case of the adiabatic approximation
 and apply them in section 6 for analysis of the evolution of the
 soliton parameters. A unique possibility is revealed to suppress
 the perturbation-induced shift of the soliton centre at the cost
 of proper matching of the soliton width and nonlinearity
 dispersion parameter. In section 7 we calculate the spectral
 density of the radiation power emitted by the MNLS DM-soliton.
 Finally, section 8 contains a summary of the paper.

 \section{Basic equations}

 We model dynamics of a subpicosecond optical pulse in a cascaded
 transmission system with periodically varying chromatic
 dispersion by the MNLS equation:
 \be\label{1.1}
 {\rm i}q_z+\frac{1}{2}D(z)q_{tt}+{\rm i}\alpha(|q|^2q)_t+|q|^2q={\rm i}Gq,
 \ee
 \[
 G=-\gamma+\left({\rm e}^{\gamma
 \ell}-1\right)\sum_{n=1}^N\delta(z-n\ell).
 \]
 Here all quantities are dimensionless: $q(z,t)$ is the complex
 envelope of the electric field $E$ normalized to the typical
 pulse power $|E|^2=P_0|q|^2$, $t$ is the retarded time normalized
 to the typical pulse width, the distance $z$ down the fibre is
 normalized to the nonlinear length $L_{nl}=(\sigma P_0)^{-1}$,
 $\sigma$ is the nonlinear coefficient. $D(z)$ is the chromatic
 dispersion and $\alpha$ measures the strength of the
 nonlinearity dispersion. The loss coefficient $\gamma$ and
 $\alpha$ are supposed to be $z$-independent and $\ell$ is the
 amplifier spacing. Taking $q(z,t)$ in the form
 \cite{Has-Kod}
 \[
 q(z,t)=u(z,t)\exp\left[\int_0^z{\rm d}z'G(z')\right],
 \]
 we transform equation (\ref{1.1}) to
 \be\label{1.2}
 {\rm i}u_z+\frac{1}{2}D(z)u_{tt}+{\rm i}\alpha
 g(z)(|u|^2u)_t+g(z)|u|^2u=0,
 \ee
 where
 \[
 g(z)=\exp\left[2\int_0^z{\rm d}z'G(z')\right].
 \]

 As usually, we consider a transmission line comprising
 periodically alternating fibre sections and lumped amplifiers.
 Each fibre section of the length $\ell$ contains a portion of a
 compensating fibre with normal dispersion $D_-<0$ and length
 $\ell_-$ and a portion of a fibre with anomalous dispersion
 $D_+>0$. Amplifiers compensate for the fibre losses over the distance
 $\ell$. Hence, the total chromatic dispersion $D(z)$ splits into
 two components: a small constant path-average (residual)
 dispersion $\delta$ and rapidly varying local dispersion
 $\Delta(z)$. The local dispersion is periodic with zero average
 $\langle \Delta\rangle=\ell^{-1}\int_0^\ell{\rm d}z\Delta(z)=0$.
 The residual dispersion within the single fibre period is given
 by $D_{{\rm res}}=[D_-\ell_-+D_+(\ell-\ell_-)]\ell^{-1}$. We take
 the amplifier spacing $\ell$ to be equal to the DM map length.

 Different dispersion scales can be associated with the above
 dispersion map scheme: the dispersion lengths $L_\pm$
 corresponding to the local chromatic dispersions $D_\pm$ and the
 dispersion length $L_{{\rm res}}$ corresponding to the residual
 dispersion. We will consider the case of $L_+\ll L_{{\rm res}}$
 and $L_{{\rm res}}\sim L_{{\rm nl}}$ which corresponds to the
 condition $\ell\ll 1$. In accordance with this condition, it is
 natural to distinguish fast and slow scales \cite{GT96} which
 are represented by the variables $\zeta=z/\ell$ and $z$,
 respectively. Therefore,
 \[
 D(z)=\delta+\frac{1}{\ell}\Delta(\zeta).
 \]
 Following Ablowitz and Biondini \cite{AB}, we seek for a solution
 $u(\zeta,z,t)$ of equation (\ref{1.2}) as a series in powers of
 $\ell$:
 \be\label{1.3}
 u(\zeta,z,t)=u^{(0)}(\zeta,z,t)+\ell u^{(1)}(\zeta,z,t)+\ldots .
 \ee
 Writing the derivative $\partial_z$ as
 $\partial_z=\ell^{-1}\partial_\zeta+\partial_z+{\cal
 O}(\ell)$, substituting the series (\ref{1.3}) in equation
 (\ref{1.2}) and equating terms with equal powers of $\ell$, we
 obtain a chain of equations. In the leading order ${\cal
 O}(\ell^{-1})$ we arrive at a linear equation
 \be\label{1.4}
 {\rm i}u_\zeta^{(0)}+\frac{1}{2}\Delta(\zeta)u_{tt}^{(0)}=0.
 \ee
 In the frequency domain (here and below we do not explicitly
 indicate the integration limits if they are infinite)
 \[
 u^{(0)}(\zeta,z,t)=\int{\rm d}\omega{\rm
 e}^{-{\rm i}\omega t}\hat u^{(0)}(\zeta,z,\omega),
 \qquad \hat u^{(0)}(\zeta,z,\omega)
 =\frac{1}{2{\pi}}\int{\rm d}t{\rm
 e}^{{\rm i}\omega t}u^{(0)}(\zeta,z,t)
 \]
 equation (\ref{1.4}) is easily solved:
 \be\label{1.5}
 \hat u^{(0)}(\zeta,z,\omega)=\hat
 U^{(0)}(z,\omega)\exp\left[-\frac{{\rm
 i}}{2}\omega^2\left(\int_0^\zeta{\rm
 d}\zeta'\Delta(\zeta')+C_0\right)\right]\equiv\hat
 U^{(0)}(z,\omega)\hat P(\zeta,\omega).
 \ee
 Here $C_0={\rm const}$. Hence, in the leading order the pulse
 evolution is mainly determined by the local dispersion
 $\Delta(\zeta)$, while the integration constant
 $\hat U^{(0)}(z,\omega)$ depends on the slow variable $z$ and is
 determined by the next-order equation:
 \be\label{1.6}
 {\rm
 i}u^{(1)}_\zeta+\frac{1}{2}\Delta(\zeta)u_{tt}^{(1)}=-\left[{\rm
 i}u_z^{(0)}+\frac{1}{2}\delta u_{tt}^{(0)}+{\rm i}\alpha
 g(z)\left(|u^{(0)}|^2u^{(0)}\right)_t+g(z)|u^{(0)}|^2u^{(0)}\right].
 \ee
 In the frequency domain, where $u^{(1)}(\zeta,z,t)=\int{\rm
 d}\omega\exp(-{\rm i}\omega t)\hat u^{(1)}(\zeta,z,\omega)$,
 equation (\ref{1.6}) has the form
 \be\label{1.7}
 {\rm i}\hat u^{(1)}_\zeta-\frac{\omega^2}{2}\Delta(\zeta)\hat u^{(1)}
 =-\hat P(\zeta,\omega)\Biggl[{\rm i}\hat
 U^{(0)}_z(z,\omega)-\frac{1}{2}\delta\omega^2\hat U^{(0)}(z,\omega)
 \ee
 \[
 +g(z)(1+\alpha\omega)\int\!\!\int{\rm d}\omega_1{\rm
 d}\omega_2\hat U^{(0)}(z,\omega+\omega_1)\hat U^{(0)}(z,\omega+\omega_2)
 \hat{\bar
 U}^{(0)}(z,\omega+\omega_1+\omega_2)T(\zeta,\omega_1\omega_2)\Biggr],
 \]
 where overbar means complex conjugation and
 \[
 T(\zeta,\kappa)=\exp\left[{\rm i}\kappa\left(\int_0^\zeta{\rm
 d}\zeta'\Delta(\zeta')+C_0\right)\right].
 \]
 The standard procedure to remove secular terms, i.e.,
 orthogonality of the right-hand side of equation (\ref{1.7}) to a
 solution of the homogeneous equation, $\int_0^1{\rm
 d}\zeta\hat{\bar P}(\zeta, \omega)\cdot({\rm r.h.s.})=0$, leads to
 the equation for $\hat U^{(0)}(z,\omega)$:
 \be\label{1.8}
 {\rm i}\hat U^{(0)}_\zeta-\frac{\delta\omega^2}{2}\hat
 U^{(0)}
 +g(z)(1+\alpha\omega)\int\!\!\int{\rm d}\omega_1{\rm
 d}\omega_2\hat U^{(0)}(z,\omega+\omega_1)\hat U^{(0)}(z,\omega+\omega_2)
 \hat{\bar
 U}^{(0)}(z,\omega+\omega_1+\omega_2)T(\omega_1\omega_2)=0,
 \ee
 \be\label{1.9}
 T(\kappa)=\int_0^1{\rm d}\zeta\exp\left[{\rm
 i}\kappa\left(\int_0^\zeta{\rm
 d}\zeta'\Delta(\zeta')+C_0\right)\right].
 \ee
 Equation (\ref{1.8}) describes the averaged dynamics of the pulse
 with all fast and large variations removed and represents a
 straightforward generalization of the Gabitov--Turitsyn equation
 \cite{GT96} when accounting for the nonlinearity dispersion.

\begin{figure}
 \begin{center}
 \includegraphics[scale=0.45]{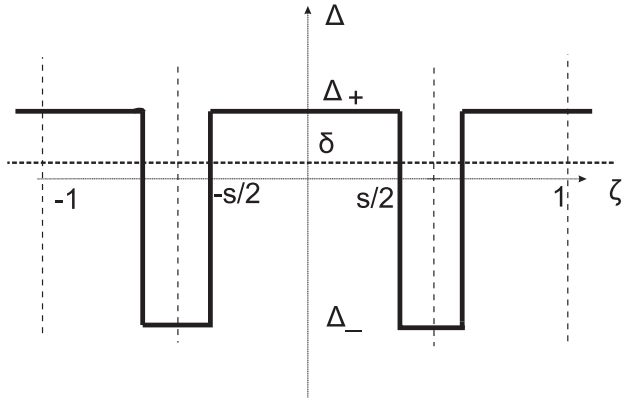}
 \caption{Two-step dispersion map with the residual dispersion $\delta$ and
 local dispersions $\Delta_\pm$.}
 \label{fig.1}
 \end{center}
 \end{figure}

 It is easy to explicitly calculate the function $T(\kappa)$
 (\ref{1.9}) for the piecewise two-step dispersion map (Figure \ref{fig.1}),
 given by
 \[
 \Delta(\zeta)=\left\{\begin{array}{ll}\Delta_+,& \quad
 \displaystyle{-\frac{s}{2}<\zeta<\frac{s}{2}},\\ \Delta_-,& \quad
 \displaystyle{\zeta\in\left(-\frac{1}{2},-\frac{s}{2}\right], \quad
 \zeta\in\left[\frac{s}{2},\frac{1}{2}\right),}\end{array}\right.
 \]
 where $s=-\Delta_-(\Delta_+-\Delta_-)^{-1}$ is determined from
 the condition $\langle\Delta\rangle=0$. As a result, we obtain
 \[
 T(\omega_1\omega_2)=\frac{\sin(\mu\omega_1\omega_2)}{\mu\omega_1\omega_2}.
 \]
 Here $\mu=(1/4)[\Delta_+s-\Delta_-(1-s)]$ measures the normalized
 map strength.

 In what follows we will consider the case of small $\mu$ that
 corresponds to the limit of weak and perhaps moderate dispersion
 management. It gives
 \be\label{1.9a}
 T(\omega_1\omega_2)=1-\frac{1}{6}\mu^2\omega_1^2\omega_2^2,
 \qquad \mu^2\ll1.
 \ee
 As a result, equation (\ref{1.8}) takes the form
 \be\label{1.10}
 {\rm i}\hat U^{(0)}_\zeta-\frac{\delta\omega^2}{2}\hat
 U^{(0)}
 +g(z)(1+\alpha\omega)\int\!\!\int{\rm d}\omega_1{\rm
 d}\omega_2\hat U^{(0)}(z,\omega+\omega_1)\hat U^{(0)}(z,\omega+\omega_2)
 \hat{\bar
 U}^{(0)}(z,\omega+\omega_1+\omega_2)
 \ee
 \[
 =\frac{1}{6}\mu^2g(z)(1+\alpha\omega)\int\!\!\int{\rm d}\omega_1{\rm
 d}\omega_2\hat U^{(0)}(z,\omega+\omega_1)\hat U^{(0)}(z,\omega+\omega_2)
 \hat{\bar
 U}^{(0)}(z,\omega+\omega_1+\omega_2).
 \]
 Before we proceed further and come back to the temporal domain,
 we scale the coordinate $z$ and the function $\hat U^{(0)}$: $z=z'/\delta$,
 $\hat U=(\delta/g)^{1/2}\hat U'$. Performing the inverse Fourier
 transform and omitting primes, we arrive at the perturbed MNLS
 equation:
 \be\label{1.11}
 {\rm i}u_z+\frac{1}{2}u_{tt}+{\rm i}\alpha
 (|u|^2u)_t+|u|^2u=\frac{1}{6}\mu^2(r+{\rm i}\alpha r_t)
 \ee
 with the perturbation
 \be\label{1.12}
 r(z,t)=u^2\bar u_{tttt}+4uu_t\bar u_{ttt}+2(uu_{tt}+2u_t^2)\bar
 u_{tt}+4u_tu_{tt}\bar u_t+u_{tt}^2\bar u.
 \ee

 It should be noted that the form (\ref{1.12}) of the perturbation
 for the standard NLS-based DM soliton has been previously obtained by
 Lakoba and Pelinovsky \cite{LP} and Boscolo {\it et al.}
 \cite{Sonia}. A contribution of the nonlinearity dispersion
 manifests itself by the $\alpha$-dependent terms both in the left
 and right hand sides of equation (\ref{1.11}). In the next two
 Sections we will derive soliton solution of the MNLS equation and
 develop a perturbation theory to apply it to equation
 (\ref{1.11}).

 \section{Integrable MNLS equation}

 The unperturbed MNLS equation
 \be\label{2.1}
 {\rm i}u_z+\frac{1}{2}u_{tt}+{\rm i}\alpha
 (|u|^2u)_t+|u|^2u=0
 \ee
 admits the Lax representation $\mathrm{\bf A}_z-\mathrm{\bf B}_t+[\mathrm{\bf
 A},\mathrm{\bf B}]=\mathrm{\bf 0}$ with the matrices $\mathrm{\bf A}$ and $\mathrm{\bf
 B}$ of the form
 \begin{eqnarray}\label{2.2}
 \mathrm{\bf A}&=&\Lambda(k)\bm\sigma_3+2{\rm i}k\mathrm{\bf Q},
 \qquad \mathrm{\bf Q}=\left(\begin{array}{cc} 0
 & u\\ \bar u & 0\end{array}\right),\\
 \mathrm{\bf B}&=&\Omega(k){\bm\sigma}_3-2k\Lambda(k)\mathrm{\bf
 Q}+2{\rm i}k^2\mathrm{\bf Q}^2{\bm\sigma}_3+k\mathrm{\bf Q}_t{\bm
 \sigma}_3-2{\rm i}\alpha k\mathrm{\bf Q}^3. \nonumber
 \end{eqnarray}
 Here \be\label{2.3} \Lambda(k)=-\frac{{\rm i}}{2\alpha}(4k^2-1),
 \qquad\Omega(k)={\rm i}\Lambda^2(k)=-\frac{{\rm
 i}}{4\alpha^2}(4k^2-1)^2, \ee $k$ is a spectral parameter and
 ${\bm\sigma}_3$ is the Pauli matrix. In the limit $\alpha\to 0$
 equation (\ref{2.1}) goes to the NLS equation. As regards the
 matrices $\mathrm{\bf A}$ and $\mathrm{\bf B}$, the NLS limit is
 less trivial and is given by representing $k$ in the form
 \cite{Kut}
 \be\label{2.4}
 k=\frac{1}{2}(1+\alpha\lambda)+{\cal O}(\alpha^2).
 \ee
 Then the limit $\alpha\to 0$ transforms the matrices $\mathrm{\bf
 A}$ and $\mathrm{\bf B}$ into the standard Lax matrices of the
 NLS equation with the spectral parameter $\lambda$.

 \subsection{Jost solutions}

 The associated linear problem for the MNLS equation is written as
 \be\label{2.5}
 \mathrm{\bf J}_t=\Lambda(k)[{\bm\sigma}_3,\mathrm{\bf J}]
 +2{\rm i}k\mathrm{\bf Q}
 \mathrm{\bf J}
 \ee
 for the matrix-valued function $\mathrm{\bf J}(t,k)$ (we will omit
 the evolution variable $z$, unless evolution aspects are
 concerned). Solutions of the spectral equation (\ref{2.5}) obey
 two important symmetries, namely, the involution
 \be\label{2.7}
 \mathrm{\bf J}^\dag(t,k)=\mathrm{\bf J}^{-1}(t,k)
 \ee
 with $\dag$ denoting the Hermitian conjugation, and the parity
 \be\label{2.7}
 \mathrm{\bf J}(t,k)={\bm\sigma}_3\mathrm{\bf
 J}(t,-k){\bm\sigma}_3.
 \ee

 Matrix Jost solutions $\mathrm{\bf J}_\pm$ of equation
 (\ref{2.5}) obey the asymptotic conditions $\mathrm{\bf
 J}_\pm\to{\bm 1}$ as $t\to\pm\infty$. Since ${\mathrm {tr}}\mathrm{\bf
 A}=0$, we have $\det\mathrm{\bf J}_\pm=1$ for all $t$. Being
 solutions of the first-order equation (\ref{2.5}), the Jost
 functions are not independent. Indeed, they are interconnected by
 the scattering matrix $\mathrm{\bf S}$:
 \be\label{2.8}
 \mathrm{\bf J}_-=\mathrm{\bf J}_+\mathrm{\bf E}\mathrm{\bf S}\mathrm{\bf
 E}^{-1}, \qquad \mathrm{\bf S}(k)=\left(\begin{array}{rr}a &
 -\bar b\\ b& \bar a\end{array}\right),
 \qquad \det\mathrm{\bf S}=1, \qquad \mathrm{\bf
 E}=\exp\left[\Lambda(k)t{\bm\sigma}_3\right].
 \ee
 For the subsequent analysis, analytic properties of the Jost
 solutions are of primary importance.

 \subsection{Analytic solutions}

 Rewriting the spectral equation (\ref{2.5}) for the Jost
 functions with the corresponding boundary conditions in the form
 of the Volterra integral equations, we obtain for the first
 column $\mathrm{\bf J}_-^{[1]}$ of the Jost matrix $\mathrm{\bf
 J}_-$:
 \begin{eqnarray}\label{2.9}
 J_{-11}(t,k)&=&1+2{\rm i}k\int_{-\infty}^t{\rm
 d}t'u(t')J_{-21}(t',k) \\
 J_{-21}(t,k)&=&2{\rm i}k\int_{-\infty}^t{\rm d}t'\bar
 u(t',k)J_{-11}(t',k)\exp\left[\frac{{\rm
 i}}{\alpha}(4k^2-1)(t-t')\right].\nonumber
 \end{eqnarray}
 It follows from the integrand in (\ref{2.9}) that the column $\mathrm{\bf
 J}_-$ is analytic in the domain ${\mathbb C}_+=\{k, \; \mathrm
 {Im}k^2>0\}$, i.e., in the first and third quadrants of the
 complex $k$-plane, and continuous for $\mathrm{
 Im}k^2=0$, i.e., on the real and imaginary axes. The same result
 follows for the second column $\mathrm{\bf J}_+^{[2]}$ of the
 Jost matrix $\mathrm{\bf J}_+$. Therefore, the matrix function
 \be\label{2.10}
 \bm\Phi_+=\left(\mathrm{\bf J}_-^{[1]},\;\mathrm{\bf
 J}_+^{[2]}\right)
 \ee
 solves the spectral problem (\ref{2.5}) and analytic as a whole
 in ${\mathbb C}_+$. The analytic function $\mathrm{\bf\Phi}_+$ can
 be expressed in terms of the Jost functions and some elements of
 the scattering matrix:
 \be\label{2.11}
 \bm\Phi_+=\mathrm{\bf J}_+\mathrm{\bf E}\mathrm{\bf S}_+
 \mathrm{\bf E}^{-1}=\mathrm{\bf J}_-\mathrm{\bf E}\mathrm{\bf
 S}_-\mathrm{\bf E}^{-1},
 \ee
 \[
 \mathrm{\bf S}_+=\left(\begin{array}{cc}a & 0\\b &
 1\end{array}\right), \qquad \mathrm{\bf S}_-=\left(\begin{array}{cc}1 & \bar b
 \\0 &
 a\end{array}\right), \qquad \mathrm{\bf S}_+=\mathrm{\bf S}\mathrm{\bf
 S}_-.
 \]
 Equation (\ref{2.11}) yields
 \be\label{2.12}
 \det\bm\Phi_+(t,k)=a(k).
 \ee
 The function $\bm\Phi_+$ can be expanded in an asymptotic series
 in $k^{-1}$:
 \be\label{2.13}
 \bm\Phi_+(t,k)=\bm\Phi_+^{(0)}(t)+k^{-1}\bm\Phi_+^{(1)}(t)+{\cal
 O}(k^{-2}).
 \ee
 In virtue of the parity property (\ref{2.7}), expansion
 coefficients possess a definite structure. Namely, matrices
 $\bm\Phi_+^{(2n)}(t)$, $n=0,1,,\ldots$, are diagonal, while
 $\bm\Phi_+^{(2n+1)}(t)$  are off-diagonal. Substituting the
 expansion (\ref{2.13}) into the spectral equation (\ref{2.5})
 yields the formula for the reconstruction of the potential $\mathrm{\bf
 Q}$:
 \be\label{2.14}
 \mathrm{\bf Q}=\frac{2}{\alpha}\bm\sigma_3\bm\Phi_+^{(1)}\left(\bm\Phi_+^{(0)}
 \right)^{-1},
 \ee
 as well as the equation for the leading-order term
 $\bm\Phi_+^{(0)}$:
 \be\label{2.15}
 \bm\Phi_{+t}^{(0)}=-{\rm i}\alpha\bm\sigma_3\mathrm{\bf
 Q}^2\bm\Phi_+^{(0)}.
 \ee
 Similarly we introduce a matrix function
 $\bm\Phi_-^{-1}(t,k)$ made up from the rows $\left(\mathrm{\bf
 J}_\pm^{-1}\right)_{[j]}$ of the matrices $\mathrm{\bf
 J}_\pm^{-1}$:
 \[
 \bm\Phi_-^{-1}=\left(\begin{array}{l}\left(\mathrm{\bf
 J}_-^{-1}\right)_{[1]} \\ \left(\mathrm{\bf
 J}_+^{-1}\right)_{[2]}\end{array}\right).
 \]
 This matrix is a solution of the adjoint spectral problem and
 analytic in the domain $\mathbb{C}_-=\{k,\;\mathrm{Im}k^2<0\}$,
 i.e., in the second and fourth quadrants of the $k$-plane, and
 continuous for $\mathrm{Im}\;k^2=0$. Besides,
 \be\label{2.16}
 \bm\Phi_-^{-1}=\mathrm{\bf E}\mathrm{\bf R}_+\mathrm{\bf E}^{-1}\mathrm{\bf J}
 _+^{-1}=\mathrm{\bf E}\mathrm{\bf R}_-\mathrm{\bf E}^{-1}\mathrm{\bf J}
 _-^{-1},
 \ee
 \[
 \mathrm{\bf R}_+=\left(\begin{array}{cc}\bar a & \bar b\\ 0 &
 1\end{array}\right), \qquad \mathrm{\bf R}_-=\left(\begin{array}{cc}
 1 & 0\\ b& a\end{array}\right), \qquad \mathrm{\bf R}_+\mathrm{\bf
 S}=\mathrm{\bf R}_-.
 \]
 Hence, $\det\bm\Phi_-^{-1}(t,k)=\bar a(k)$ and the involution
 \be\label{2.17}
 \bm\Phi_-^{-1}(t,\bar k)=\bm\Phi_+^\dag(t,k)
 \ee
 takes place as well.

 \subsection{The Riemann--Hilbert problem}

 It follows from equation (\ref{2.14}) that it is the analytic
 function $\bm\Phi_+$ that determines a solution of the MNLS
 equation. On the other hand, the availability of two matrix
 functions $\bm\Phi_+$ and $\bm\Phi_-^{-1}$ which are analytic in
 the complementary domains $\mathbb{C}_\pm$ of the $k$-plane and
 continuous on the common contour $\mathrm{Im}\;k^2=0$ permits to
 pose the matrix RH problem for these functions:
 \be\label{2.18}
 \bm\Phi_-^{-1}(t,k)\bm\Phi_+(t,k)=\mathrm{\bf E}\mathrm{\bf G}(k)
 \mathrm{\bf E}^{-1},
 \ee
 \be\label{2.18a}
 \mathrm{\bf G}(k)=\mathrm{\bf R}_+\mathrm{\bf
 S}_+=\left(\begin{array}{cc}1 & \bar b\\ b &
 1\end{array}\right),
 \qquad k\in\mathrm{Im}\;k^2=0.
 \ee
 Equation (\ref{2.18}) is easily obtained from equations (\ref{2.11})
 and (\ref{2.16}). In other words, the RH problem (\ref{2.18})
 poses a problem of analytic factorisation of the non-degenerate
 matrix function $\mathrm{\bf G}(k)$ specified on the contour
 $\mathrm{Im}\;k^2=0$, in a product of two matrix functions
 $\bm\Phi_+$ and $\bm\Phi_-^{-1}$ analytic in the corresponding
 domains of the $k$-plane. Hence, solving the MNLS equation is equivalent
 to solution of the RH problem (\ref{2.18}). The normalization
 condition necessary to provide the uniqueness of solution of the
 RH problem follows from equation (\ref{2.13}):
 \be\label{2.19}
 \bm\Phi_+(t,k)\rightarrow\bm\Phi_+^{(0)}(k) \quad \mathrm{at}
 \quad k\to \infty.
 \ee
 Hence, we obtain the RH problem with the non-standard
 normalization ($\bm\Phi_+^{(0)}$ instead of the unit matrix), as
 distinct from the NLS case.

 In general, the matrices $\bm\Phi_+$ and $\bm\Phi_-^{-1}$ can
 have zeros $k_j$ and $\bar k_\ell$ in the corresponding analyticity
 domains: $\det\bm\Phi_+(k_j)=0$, $k_j\in\mathbb{C}_+$ and
 $\det\bm\Phi_-^{-1}(\bar k_\ell)=0$, $\bar
 k_\ell\in\mathbb{C}_-$. In virtue of the parity property
 (\ref{2.7}) zeros appear in pairs, $\pm k_j$ and $\pm \bar
 k_\ell$. The involution property (\ref{2.17}) guarantees equal
 number $\cal{N}$ of zeros which we consider as simple. They are
 zeros of the RH problem that determine soliton solutions of the
 MNLS equation.

 \subsection{Regularization of the RH problem}

 We will solve the RH problem with zeros by means of its
 regularization, i.e., by extracting rational factors from
 $\bm\Phi_\pm$ that are responsible for the appearance of zeros.
 Namely, let us multiply $\bm\Phi_+(t,k)$ having a single zero
 $k_j$,
 by a rational function $\bm\Xi_j^{-1}(t,k)$ which has a simple pole
 in $k_j$. Then the product $\bm\Phi_+(t,k)\bm\Xi_j^{-1}(t,k)$
 will be regular in $k_j$. Since $\det\bm\Phi_+(k_j)=0$, there
 exists an eigenvector $|\chi_j\rangle$ with zero eigenvalue, $\bm\Phi_+(k_j)
 |\chi_j\rangle=0$. We take $\bm\Xi_j^{-1}$ in the form
 \be\label{2.20}
 \bm\Xi_j^{-1}(t,k)={\bm 1}+\frac{k_j-\bar k_j}{k-k_j}\mathrm{\bf P}_j(t),
 \qquad
 \mathrm{\bf P}_j(t)=\frac{|\chi_j\rangle\langle\chi_j|}{\langle\chi_j|\chi_j\rangle},
 \qquad \langle\chi_j|=|\chi_j\rangle^\dag,
 \ee
 where $\mathrm{\bf P}_j$ is a projector of rank 1,
 $\mathrm{\bf P}_j^2=\mathrm{\bf P}_j$. It is easy to
 see that $\det\Xi_j^{-1}=(k_j-\bar k_j)(k-k_j)^{-1}$, as should
 be to have a pole in $k_j$. The regularization in the point
 $-k_j$ is evidently given by
 \[
 \bm\Xi_{-j}^{-1}(t,k)={\bm 1}-\frac{k_j-\bar
 k_j}{k+k_j}\mathrm{\bf P}_{-j}(t),
 \]
 where, due to the parity property (\ref{2.7}),
 $|\chi_{-j}\rangle=\bm\sigma_3|\chi_j\rangle$ and
 $\mathrm{\bf P}_{-j}=\bm\sigma_3\mathrm{\bf P}_j\bm\sigma_3$. As a result,
 the matrix
 function $\bm\Phi_+\bm\Xi_j^{-1}\bm\Xi_{-j}^{-1}$ is regular in
 $\pm k_j$. Similarly, the matrix
 $\bm\Xi_{-\ell}\bm\Xi_\ell\bm\Phi_-^{-1}$ has no zeros in $\pm\bar
 k_\ell$, where
 \be\label{2.21}
 \bm\Xi_\ell={\bm 1}-\frac{k_\ell-\bar k_\ell}{k-\bar
 k_\ell}\mathrm{\bf P}_\ell(t), \qquad \bm\Xi_{-\ell}(t,k)={\bm
 1}+\frac{k_\ell-\bar k_\ell}{k+\bar k_\ell}\mathrm{\bf P}_{-\ell}(t).
 \ee
 Regularising all $4\cal{N}$ zeros of the RH problem, we represent
 the matrices $\bm\Phi_\pm$ as
 \be\label{2.22}
 \bm\Phi_\pm=\bm\phi_\pm\bm\Gamma, \qquad
 \bm\Gamma=\bm\Xi_{-\cal{N}}\bm\Xi_{\cal{N}}\cdots\bm\Xi_{-1}\bm\Xi_1.
 \ee
 Here the matrices $\bm\phi_\pm(t,k)$ solve the regular RH
 problem, i.e., without zeros,
 \be\label{2.23}
 \bm\phi_-^{-1}\bm\phi_+=\bm\Gamma\mathrm{\bf E}\mathrm{\bf G}
 \mathrm{\bf E}^{-1}\bm\Gamma^{-1},
 \ee
 while the rational matrix $\bm\Gamma(t,k)$ accumulates all zeros
 of the RH problem. Evidently,
 $\det\bm\Gamma=\prod_{j=1}^{\cal{N}}(k-k_j)(k-\bar k_j)^{-1}$. It
 follows from the structure of the rational multipliers $\bm\Xi$
 that the asymptotic expansion for $\bm\Gamma$ starts with the
 unit matrix:
 \be\label{2.24}
 \bm\Gamma(t,k)={\bf 1}+k^{-1}\bm\Gamma^{(1)}(t)+{\cal O}(k^{-2}).
 \ee
 Besides, the matrix $\bm\Gamma$ hereditates the involution
 property $\bm\Gamma^\dag(t,k)=\bm\Gamma^{-1}(t,\bar k)$.

 For the unperturbed MNLS equation, we can take solutions
 $\bm\phi_\pm$ of the regular RH problem (\ref{2.23}) being
 $k$-independent. Hence, comparing the asymptotic expansions
 (\ref{2.13}) and (\ref{2.24}), we obtain that the solution of the
 regular RH problem is given by the leading-order term
 $\bm\Phi_+^{(0)}$:
 \be\label{2.25}
 \bm\phi_+(t)=\bm\Phi_+^{(0)}(t).
 \ee

 In practice, a successive application of the elementary
 multipliers $\bm\Xi_j$ to regularize the RH problem would not be
 an optimal way. It is more convenient to decompose the product of
 $\bm\Xi_j$'s (\ref{2.22}) into simple fractions. After some
 algebra \cite{Kawata,Vas} we obtain
 \begin{eqnarray}\label{2.26}
 \bm\Gamma(t,k)&=&{\bf 1}-\sum_{m,n=1}^{2{\cal
 N}}(k-\bar\kappa_n)|m\rangle\left(D^{-1}\right)_{mn}\langle n|,\\
 \bm\Gamma^{-1}(t,k)&=&{\bf 1}+\sum_{m,n=1}^{2{\cal
 N}}(k-\kappa_m)|m\rangle\left(D^{-1}\right)_{mn}\langle
 n|\nonumber
 \end{eqnarray}
 with new vectors $|m\rangle$ (in the same way as
 $|\chi_\pm\rangle$, the vectors $|m\rangle$ still solve the
 equation $\bm\Phi_+(\kappa_m)|m\rangle=0$) and matrix $\mathrm{\bf
 D}$ with the entries
 \be\label{2.27}
 \mathrm{\bf D}_{nm}=\frac{\langle
 n|m\rangle}{\kappa_m-\bar\kappa_ n}.
 \ee
 We adopt the following enumeration of zeros:
 $\kappa_m=\{k_1,-k_1,k_2,-k_2,\ldots,k_{\cal N},-k_{\cal N}\}$.

 To determine coordinate dependence of $|m\rangle$, we turn to the
 equation $\bm\Phi_+(\kappa_m)|m\rangle=0$. Differentiating it in
 $t$ and accounting equation (\ref{2.5}) yields
 $|m\rangle_t=\Lambda(\kappa_m)\bm\sigma_3|m\rangle$. In the same
 way, differentiation in $z$ together with the evolution equation
 \be\label{2.28}
 \bm\Phi_{+z}=\mathrm{\bf
 B}\bm\Phi_+-\Omega(k)\bm\sigma_3\bm\Phi_+
 \ee
 gives $|m\rangle_z=\Omega(\kappa_m)\bm\sigma_3|m\rangle$.
 Therefore, the coordinate dependence of the vector $|m\rangle$
 has the very simple form:
 \be\label{2.29}
 |m\rangle=\exp\left\{\left[\Lambda(\kappa_m)t+\Omega(\kappa_m)z\right]
 \bm\sigma_3\right\}|m_0\rangle,
 \ee
 where $|m_0\rangle=(p_1,p_2)^T$ is a constant vector. The parity
 property (\ref{2.7}) gives $|2m\rangle=\bm\sigma_3|2m-1\rangle$.

 Zeros $\kappa_m$ and vectors $|m\rangle$ comprise the discrete
 data of the RH problem that determine the soliton content of a
 solution to the MNLS equation. The continuous RH data is
 characterized by the matrix $\mathrm{\bf G}(k)$,
 $k\in\mathrm{Im}\;k^2$ (or by the scalar function $b(k)$ entering
 $\mathrm{\bf G}(k)$ (\ref{2.18a})) and is responsible for the
 description of radiation. The evolution equation for $\mathrm{\bf
 G}(k)$,
 \be\label{2.33}
 \mathrm{\bf G}_z=\Omega(k)[\bm\sigma_3,\;\mathrm{\bf G}],
 \ee
 easily follows from equations (\ref{2.18}) and (\ref{2.28}).

 \subsection{Reconstruction of the potential}

 The rational function $\bm\Gamma(t,k)$ plays the key role in both
 finding solitons and describing its perturbed dynamics in the
 adiabatic approximation. Indeed, equations (\ref{2.5}),
 (\ref{2.14}) and (\ref{2.22}) give
 \[
 \bm\Gamma_t=\left[\left(\bm\Phi_+^{(0)}\right)^{-1}\bm\Phi_+\right]_t
 ={\rm i}\alpha\bm\sigma_3\mathrm{\bf
 Q}^2\left(\bm\Phi_+^{(0)}\right)^{-1}
 \bm\Phi_++\left(\bm\Phi_+^{(0)}\right)^{-1}\left(\Lambda(k)[\bm\sigma_3,\;
 \bm\Phi_+]+2{\rm i}k\mathrm{\bf Q}\bm\Phi_+\right).
 \]
 Hence, the equation for
 $\bm\Gamma_0^{-1}\equiv\left[\bm\Gamma(k=0)\right]^{-1}$,
 \be\label{2.30}
 \left(\bm\Gamma_0^{-1}\right)_t=-{\rm i}\alpha\bm\sigma_3\mathrm{\bf
 Q}^2\bm\Gamma_0^{-1},
 \ee
 coincides with equation (\ref{2.15}) for $\bm\Phi_+^{(0)}$.
 Identifying both functions,
 \be\label{2.31}
 \bm\Gamma_0^{-1}=\bm\Phi_+^{(0)},
 \ee
 we derive from (\ref{2.14}) and (\ref{2.25}) a formula for
 soliton solutions in terms of $\bm\Gamma$:
 \be\label{2.32}
 \mathrm{\bf
 Q}=\frac{2}{\alpha}\bm\sigma_3\bm\Gamma_0^{-1}\bm\Gamma^{(1)}\bm\Gamma_0.
 \ee
 Here $\bm\Gamma^{(1)}$ is the coefficient in the series
 (\ref{2.24}). Note that equation (\ref{2.32}) cannot be applied
 to consider radiation emitted by a perturbed soliton.

 \subsection{MNLS soliton}

 Now we apply the formalism developed in the preceding Section to
 get a single soliton solution of MNLS. The RH data are purely
 discrete: ${\cal N}=1$, $b(k)=0$ (or $\mathrm{\bf G}(k)={\bf
 1})$, $\bm\phi_+=\bm\phi_-$. As it follows from (\ref{2.32}), to
 obtain the soliton, we should know the rational function
 $\bm\Gamma$ which in turn is completely determined by zeros
 $\kappa_1$ and $\kappa_2$ (i.e., by $\pm k_1$), as well as by the
 vectors $|1\rangle$ and $|2\rangle=\bm\sigma_3|1\rangle$. From
 (\ref{2.3}) and (\ref{2.29}) we obtain the vector $|1\rangle$
 explicitly:
 \be\label{2.34}
 |1\rangle={\rm e}^{(1/2)(\tau_0+{\rm i}\psi_0)}\left(\begin{array}{l}
 {\rm e}^{(1/2)(\tau+{\rm i}\psi)}\\ {\rm e}^{(-1/2)(\tau+{\rm
 i}\psi)}
 \end{array}\right).
 \ee
 Here $\tau$ and $\psi$ are natural soliton coordinates related
 linearly to $t$ and $z$:
 \begin{eqnarray}\label{2.35}
 \tau&=&w(t-\tilde t\,), \qquad \tilde t=vz-\tau_0/w,
 \\
 \psi&=&\frac{v}{w}\tau+\widetilde\psi, \qquad
 \widetilde\psi=\frac{1}{2}(v^2+w^2)z+\widetilde\psi_0, \qquad
 \widetilde\psi_0=-\frac{v}{w}\tau_0+\psi_0.\nonumber
 \end{eqnarray}
 The real parameters $v$ and $w$ are constants which are
 determined by the zeros $k_1$ and $\bar k_1$ and will be identified
 with the soliton
 velocity and inverse width, respectively:
 \be\label{2.36}
 v=\frac{1}{\alpha}\left[1-2(k_1^2+\bar k_1^2)\right], \qquad
 w=\frac{2}{{\rm i}\alpha}(k_1^2-\bar k_1^2);
 \ee
 $\tau_0$ and $\psi_0$ are real parameters which are given by the
 components of the constant vector $|1_0\rangle$ (\ref{2.29}),
 $p_1/p_2=\exp(\tau_0+{\rm i}\psi_0)$. Note for the future use
 that the constant soliton parameters acquire in general a slow
 $z$-dependence in the presence of perturbation. This results in a
 modification of equations (\ref{2.35}):
 \be\label{2.37}
 \tilde t=\frac{1}{w}\left(\int_0^z{\rm
 d}z'v(z')w(z')-\tau_0\right),\qquad \widetilde\psi=v\tilde
 t-\frac{1}{2}\int_0^z{\rm
 d}z'\left(v^2(z')-w^2(z')\right)+\psi_0.
 \ee

 Having explicitly the vector $|1\rangle$, we find from
 (\ref{2.26}) the matrix $\bm\Gamma$:
 \be\label{2.38}
 \bm\Gamma={\bf 1}-\frac{\widetilde{\mathrm{\bf D}}_-}{k-\bar k_1}-
 \frac{\widetilde{\mathrm{\bf D}}_+}{k+\bar k_1}, \qquad
 \bm\Gamma^{-1}={\bf 1}+\frac{\mathrm{\bf D}_-}{k-k_1}+
 \frac{\mathrm{\bf D}_+}{k+k_1},
 \ee
 where
 \[
 \mathrm{\bf D}_-=\frac{k_1^2-\bar
 k_1^2}{2}\left(\begin{array}{cc} \displaystyle{\frac{{\rm e}^\tau}{c_+}} &
 \displaystyle{\frac{{\rm e}^{{\rm i}\psi}}{c_-}} \\ \displaystyle{
 \frac{{\rm e}^{{-\rm
 i}\psi}}{c_+}}& \displaystyle{\frac{{\rm e}^{-\tau}}{c_-}}\end{array}\right), \qquad
 \widetilde{\mathrm{\bf D}}_-=\frac{k_1^2-\bar
 k_1^2}{2}\left(\begin{array}{cc} \displaystyle{\frac{{\rm e}^\tau}{c_-}} &
 \displaystyle{\frac{{\rm e}^{{\rm i}\psi}}{c_-}} \\
 \displaystyle{\frac{{\rm e}^{{-\rm
 i}\psi}}{c_+}}& \displaystyle{\frac{{\rm e}^{-\tau}}{c_+}}\end{array}\right),
 \]
 \[
 \mathrm{\bf D}_+=-\bm\sigma_3\mathrm{\bf D}_-\bm\sigma_3, \qquad
 \widetilde{\mathrm{\bf D}}_+=-\bm\sigma_3\widetilde{\mathrm{\bf
 D}}_-\bm\sigma_3,
 \]
 and the functions $c_\pm$ are given by
 \be\label{2.39}
 c_\pm=k_1{\rm e}^{\pm\tau}+\bar k_1{\rm e}^{\mp\tau}.
 \ee
 Substituting this $\bm\Gamma$ in (\ref{2.32}) gives a simple
 expression for the MNLS soliton:
 \be\label{2.40}
 u_s=\frac{w}{{\rm i}}\frac{c_-}{c_+^2}{\rm e}^{{\rm i}\psi}.
 \ee

 This soliton solution, like that of the NLS equation, depends on
 four real parameters: velocity $v$, inverse width $w$, initial
 position $\tau_0$ and initial phase $\psi_0$. Nevertheless, its
 properties essentially differ from those of the NLS soliton.
 Indeed, the soliton (\ref{2.40}) has no habitual sech-like shape
 though its envelope is close to the hyperbolic secant. The square
 of module has the form
 \[
 |u_s|^2=\frac{1}{2}w^2\left[1-\alpha v+\sqrt{(1-\alpha
 v)^2+\alpha^2w^2}\cosh(2w(\tau-vz))\right]^{-1}.
 \]
 Hence, the envelope $|u_s|$ moves holding its shape and there is
 no any self-steepening, contrary to the behavior of the NLS
 soliton under the action of the nonlinearity dispersion. The
 reason lies in the phase properties of the MNLS soliton. Namely,
 we can write $u_s$ as $|u_s|\exp({\rm i}\psi+{\rm i}\psi_{nl})$,
 where
 \[
 \psi_{nl}=\arctan\left(\tan\theta\tanh\tau\frac{3-\tan^2\theta\tanh^2\tau}
 {1-3\tan^2\theta\tanh^2\tau}\right), \qquad k_1=|k_1|{\rm
 e}^{{\rm i}\theta}.
 \]
 We see that the MNLS soliton is characterized by a highly
 nonlinear phase and hence is intrinsically chirped. It is this
 nonlinear phase that prevents the MNLS soliton from distortion of
 its shape, as opposite to the NLS soliton whose linear phase
 cannot withstand  the self-steepening.
 Further,
 because $w\sim\alpha^{-1}$, the soliton (\ref{2.40}) contains the
 nonlinearity dispersion parameter $\alpha$ in the denominator.
 Thereby, the soliton (\ref{2.40}) is non-perturbative in $\alpha$
 and cannot be obtained in the framework of the NLS equation with
 $\alpha$-dependent perturbation. At the same time, in accordance
 with the limit procedure (\ref{2.4}), the MNLS soliton reproduces the
 standard NLS soliton in the limit
 $\alpha\to 0$. Besides, the optical energy
 \be\label{2.41}
 {\cal E}=\int{\rm d}\tau|u_s|^2=\frac{4}{\alpha}\theta
 \ee
 being the invariant of the MNLS equation, has the upper limit
 $2\pi/\alpha$ because $\theta<\pi/2$.

 \section{Perturbation theory for the MNLS soliton}

 In this Section we perform a general analysis of the perturbed
 MNLS equation
 \be\label{3.1}
 {\rm i}u_z+\frac{1}{2}u_{tt}+{\rm
 i}\alpha(|u|^2u)_t+|u|^2u=\epsilon R.
 \ee
 Here $R$ determines a functional form of a perturbation (e.g.,
 equation (\ref{1.12})), $\epsilon$ is a small parameter. In
 general, a perturbation forces the RH data to slowly evolve in
 $z$. Indeed, a perturbation causes a variation $\delta\mathrm{\bf
 Q}$ of the potential entering the spectral equation (\ref{2.5})
 and hence a variation of the Jost solutions:
 \be\label{3.2}
 \delta\mathrm{\bf J}_{\pm t}=\Lambda(k)[\sigma_3,\,\delta\mathrm{\bf
 J}_\pm]+2{\rm i}k(\delta\mathrm{\bf Q}\mathrm{\bf
 J}_\pm+\mathrm{\bf Q}\delta\mathrm{\bf J}_\pm).
 \ee
 Solving (\ref{3.2}) gives
 \be\label{3.3}
 \delta\mathrm{\bf J}_\pm=2{\rm i}k\mathrm{\bf
 J}_\pm\mathrm{\bf E}\left(\int_{\pm\infty}^t{\rm d}t'\mathrm{\bf
 E}^{-1}\mathrm{\bf J}_\pm^{-1}\delta\mathrm{\bf Q}\mathrm{\bf J}_\pm
 \mathrm{\bf E}\right)\mathrm{\bf E}^{-1}.
 \ee
 To distinguish between the 'integrable' and 'perturbative'
 contributions, we will assign the variational derivative
 $\delta/\delta t$ to the latter. Then, evidently,
 \[
 {\rm i}\frac{\delta\mathrm{\bf Q}}{\delta z}=\epsilon\mathrm{\bf R}, \qquad
 \mathrm{\bf R}=\left(\begin{array}{rr} 0 &\; R\\ -\bar R&\;0 \end{array}\right).
 \]
 By means of equations (\ref{2.8}), (\ref{2.11}) and (\ref{3.3}) we
 find a variation of the scattering matrix:
 \[
 \frac{\delta\mathrm{\bf S}}{\delta z}=2\epsilon k\mathrm{\bf S}_+
 \left(\int{\rm d}t\mathrm{\bf E}^{-1}\bm\Phi_+^{-1}\mathrm{\bf R}
 \bm\Phi_+\mathrm{\bf E}\right)\mathrm{\bf S}_-^{-1}=
 2\epsilon k\mathrm{\bf R}_+^{-1}
 \left(\int{\rm d}t\mathrm{\bf E}^{-1}\bm\Phi_-^{-1}\mathrm{\bf R}
 \bm\Phi_-\mathrm{\bf E}\right)\mathrm{\bf R}_-.
 \]
 It should be stressed that they are the analytic solutions
 $\bm\Phi_\pm$ that naturally enter this equation. Let us denote
 \be\label{3.4}
 \bm\Upsilon_\pm(t_1,t_2)=2k\int_{t_1}^{t_2}{\rm d}t\mathrm{\bf
 E}^{-1}\bm\Phi_\pm^{-1}\mathrm{\bf R}\bm\Phi_\pm\mathrm{\bf E},
 \qquad \bm\Upsilon_\pm(k)\equiv\bm\Upsilon_\pm(-\infty,\infty).
 \ee
 Then
 \be\label{3.5}
 \frac{\delta\mathrm{\bf S}}{\delta z}=\epsilon\mathrm{\bf S}_+
 \bm\Upsilon_+(k)\mathrm{\bf S}_-^{-1}=\epsilon\mathrm{\bf
 R}_+^{-1}
 \bm\Upsilon_-(k)\mathrm{\bf R}_-,
 \ee
 where the matrices $\mathrm{\bf S}_\pm$ and $\mathrm{\bf R}_\pm$
 have been defined in equations (\ref{2.11}) and (\ref{2.16}). The
 matrices $\bm\Upsilon_\pm$ are interrelated by the matrix $\mathrm{\bf
 G}$ entering the RH problem (\ref{2.18}):
 \be\label{3.6}
 \bm\Upsilon_-(k)=\mathrm{\bf G}\bm\Upsilon_+\mathrm{\bf G}^{-1}.
 \ee
 Finally, the variations of the analytic functions $\bm\Phi_\pm$
 follow from (\ref{2.24}) and (\ref{2.29}):
 \[
 \frac{\delta\bm\Phi_+}{\delta z}=\epsilon\bm\Phi_+\mathrm{\bf E}
 \bm\Pi_+\mathrm{\bf E}^{-1}, \qquad
 \frac{\delta\bm\Phi_-^{-1}}{\delta z}=\epsilon\mathrm{\bf E}
 \bm\Pi_-\mathrm{\bf E}^{-1}\bm\Phi_-^{-1}.
 \]
 Here the evolution functionals $\bm\Pi_\pm(t,k)$ are defined in
 terms of $\bm\Upsilon_\pm$ \cite{Val}:
 \be\label{3.6a}
 \bm\Pi_+(t,k)=\left(\begin{array}{ll}\Upsilon_{+11}(-\infty,t)
 &\;
 -\Upsilon_{+12}(t,\infty) \\ \Upsilon_{+21}(-\infty,t) &\;
 -\Upsilon_{+22}(t,\infty)\end{array}\right), \qquad
 \bm\Pi_-(t,k)=\left(\begin{array}{cc}\Upsilon_{-11}(-\infty,t)
 &\;
 \Upsilon_{-12}(-\infty,t) \\ -\Upsilon_{-21}(t,\infty) &\;
 -\Upsilon_{-22}(t,\infty)\end{array}\right).
 \ee
 The evolution functionals contain all needed information about a
 perturbation and enter the evolution equations for the perturbed
 RH data (see below). As a result, the evolution equations for
 $\bm\Phi_\pm$ gain additional terms responsible for the
 perturbation:
 \begin{eqnarray}\label{3.7}
 \bm\Phi_{+z}&=&\mathrm{\bf
 B}\bm\Phi_+-\bm\Phi_+\Omega(k)\bm\sigma_3+\epsilon\bm\Phi_+\mathrm{\bf
 E}\bm\Pi_+\mathrm{\bf E}^{-1},\\
 \bm\Phi_{-z}^{-1}&=&-\bm\Phi_-^{-1}\mathrm{\bf
 B}+\Omega(k)\bm\sigma_3\bm\Phi_-^{-1}-\epsilon\mathrm{\bf
 E}\bm\Pi_-\mathrm{\bf E}^{-1}\bm\Phi_-^{-1}.\nonumber
 \end{eqnarray}
 Then from equations (\ref{2.18}) and (\ref{3.7}) we obtain the
 evolution equation  for the perturbed continuous RH data:
 \be\label{3.8}
 \mathrm{\bf G}_z=\Omega(k)[\bm\sigma_3,\,\mathrm{\bf
 G}]+\epsilon(\mathrm{\bf G}\bm\Pi_+-\bm\Pi_-\mathrm{\bf G}),\qquad
 \mathrm{ Im}\,k^2=0.
 \ee
 Since the left-hand side of (\ref{3.8}) does not depend on $t$,
 we can consider this equation for $t\to\infty$. This gives
 simplified formulas for the evolution functionals:
 \be\label{3.9}
 \bm\Pi_+(k)=\left(\begin{array}{cc} \Upsilon_{+11}(k) &\; 0\\
 \Upsilon_{+21}(k) &\; 0\end{array}\right), \qquad
 \bm\Pi_-(k)=\left(\begin{array}{cc} \Upsilon_{-11}(k) &\;
 \Upsilon_{-12}(k)\\
 0&\;0 \end{array}\right).
 \ee

 To derive evolution equations for the discrete RH data, let us
 turn to the equation $\bm\Phi_+(k_1)|1\rangle=0$ which is valid
 irrespectively of the presence of a perturbation. Taking the total
 derivative in $z$, we get
 \[
 \left[\frac{{\rm d}}{{\rm
 d}z}\bm\Phi_+(k)\right]_{|k_1}|1\rangle+\bm\Phi_+(k_1)\frac{{\rm d}}{{\rm
 d}z}|1\rangle=0.
 \]
 Assuming a perturbation-induced $z$-dependence of $k_1$ yields
 \[
 \frac{{\rm d}}{{\rm
 d}z}\bm\Phi_+(k)=\mathrm{\bf
 B}\bm\Phi_+-\Omega(k)\bm\Phi_+\bm\sigma_3+\epsilon\bm\Phi_+\mathrm{\bf
 E}\bm\Pi_+\mathrm{\bf E}^{-1}+k_z\frac{\partial}{\partial
 k}\bm\Phi_+(k).
 \]
 Recall that the evolution functional $\bm\Pi_+(k)$ is expressed
 in terms of elements of the matrix $\bm\Upsilon_+$ (see
 (\ref{3.6a})) which in turn depends on $\bm\Phi_+^{-1}$ (see
 (\ref{3.4})). Therefore, $\bm\Upsilon_+$  has the simple pole in
 $k_1$ and hence $\bm\Pi_+(k)$ represents a meromorphic function
 with simple poles in zeros of $\bm\Phi_+$. This implies that near
 $k_1$ we can write
 \[
 \bm\Pi_+(k)=\bm\Pi_+^{({\rm reg})}(k)+\frac{1}{k-k_1}{\rm
 Res}\left[\bm\Pi_+(k),\,k_1\right],
 \]
 where $\bm\Pi_+^{({\rm reg})}(k)$ is a holomorphic part of
 $\bm\Pi_+$ in the point $k_1$. Following then the reasoning of
 \cite{AL}, we obtain that the perturbed evolution of the vector
 $|1\rangle$ is given by
 \[
 |1\rangle_z=\Omega(k_1)\bm\sigma_3|1\rangle-\epsilon\mathrm{\bf
 E}(k_1)\bm\Pi_+^{({\rm reg})}(k_1)\mathrm{\bf E}^{-1}|1\rangle.
 \]
 In terms of the $t$-independent vector
 $|\tilde1\rangle=\mathrm{\bf E}^{-1}(k_1)|1\rangle$ the above
 equation takes the form
 \be\label{3.10}
 |\tilde1\rangle_z=\Omega(k_1)\bm\sigma_3|\tilde1\rangle-\epsilon\left(\begin
 {array}{cc}\Upsilon_{+11}^{({\rm reg})}(k_1) &\; 0\\
 \Upsilon_{+21}^{({\rm reg})}(k_1) &\;
 0\end{array}\right)|\tilde1\rangle.
 \ee

 As regards the evolution equation for $k_1$, it follows by taking
 the total $z$-derivative of $\det\bm\Phi_+(k_1)=0$ and has the
 form
 \[
 k_{1z}=-\left[\frac{\partial_z\left(\det\bm\Phi_+(k_1)\right)}{
 \partial_k\left(\det\bm\Phi_+(k_1)\right)}\right]_{|k_1}.
 \]
 It has been shown (see (\ref{2.22})) that
 $\bm\Phi_+=\bm\phi_+\bm\Gamma$. Because
 $\bm\Gamma=\bm\Xi_{-1}\bm\Xi_1$, we have
 $\det\bm\Gamma=(k^2-k_1^2)(k^2-\bar k_1^2)^{-1}$. Besides,
 equations(\ref{2.25}) and (\ref{2.31}) gives
 $\bm\phi_+=\bm\Gamma^{-1}(k=0)$ and $\det\bm\phi_+=\bar
 k_1^2/k_1^2$. Hence,
 \[
 \det\bm\Phi_+(k)=\frac{\bar k_1^2}{k_1^2}\frac{k^2-k_1^2}{k^2-\bar
 k_1^2}.
 \]
 Taking into account the relation
 $\partial_z\left[\det\bm\Phi_+(k)\right]=\epsilon\left({\rm tr}\bm\Pi_+(k)
 \right)\det\bm\Phi_+(k)$
 and equation (\ref{3.9}), we eventually arrive at a simple
 evolution equation for the zero $k_1$:
 \be\label{3.11}
 k_{1z}=-\epsilon{\rm
 Res}\left[{\rm tr}\bm\Pi_+(k),k_1\right]=-\epsilon{\rm
 Res}\left[\Upsilon_{+11}(k),k_1\right].
 \ee

 Hence, equations (\ref{3.8}), (\ref{3.10}) and (\ref{3.11})
 determine perturbation-induced evolution of the RH data. These
 equations, however, cannot be directly applied because
 $\bm\Pi_\pm$ and $\bm\Upsilon_\pm$ entering them depend on
 unknown solutions $\bm\Phi_\pm$ of the spectral problem with the
 perturbed potential $\mathrm{\bf Q}$. The smallness of $\epsilon$
 allows us to develop the iterative scheme to consecutively
 account for two basic approximations: the leading-order adiabatic
 approximation and the next-order one.

 \section{Adiabatic approximation}

 In the framework of the adiabatic approximation we assume
 that the perturbed soliton does not radiate and adjusts its shape
 to the unperturbed one at the cost of slow evolution of its
 parameters. Hence, only the discrete RH data are relevant in this
 approximation. The soliton coordinates are given by equations
 (\ref{2.35}) and (\ref{2.37}). It is seen from (\ref{3.10}) and
 (\ref{3.11}) that the matrix $\bm\Upsilon_+$ completely
 determines evolution of the parameters. Because now
 $\bm\Phi_+=\bm\Gamma_0^{-1}\bm\Gamma$, we obtain in the adiabatic
 approximation
 \be\label{3.12}
 \bm\Upsilon_+(k)=2k\int{\rm d}t\mathrm{\bf E}^{-1}\bm\Gamma^{-1}
 \bm\Gamma_0\mathrm{\bf R}\bm\Gamma_0^{-1}\bm\Gamma\mathrm{\bf E}.
 \ee
 The explicit form of $\bm\Gamma$ is given in (\ref{2.38}).
 Therefore,
 \[
 {\rm Res}\left[\Upsilon_{+11}(k),k_1\right]=-\frac{{\rm
 i}}{2}\,\alpha \,k_1^2\int\frac{{\rm
 d}\tau}{c_-^2}\left(\rho(\tau)+\bar\rho(-\tau)\right){\rm
 e}^\tau,
 \]
 where $\rho(\tau)=R{\rm e}^{-{\rm i}\psi}$. Then equation
 (\ref{3.11}) gives a very simple equation
 \be\label{3.13}
 k_{1z}=\frac{{\rm i}}{2}\,\epsilon\,\alpha \,k_1^2\int\frac{{\rm d}\tau}{c_-^2}
 \,\rho_+\,{\rm e}^\tau,
 \ee
 where $\rho_\pm=\rho(\tau)\pm\bar\rho(-\tau)$. In addition, we
 can derive from (\ref{2.36}) evolution equations for the soliton
 velocity and inverse width:
 \be\label{3.14}
 v_z=-2{\rm i}\epsilon\int\frac{{\rm d}\tau}{c_-^2}\left(k_1^3{\rm
 e}^\tau-\bar k_1^3{\rm e}^{-\tau}\right)\rho_+(\tau),\qquad
 w_z=2\epsilon\int\frac{{\rm d}\tau}{c_-^2}\left(k_1^3{\rm
 e}^\tau+\bar k_1^3{\rm e}^{-\tau}\right)\rho_+(\tau).
 \ee

 Somewhat more cumbersome calculation is needed to derive the
 evolution equation for the vector $|\tilde 1\rangle$. In
 accordance with (\ref{3.10}) we need know $\bm\Upsilon_+^{{\rm
 (reg)}}(k_1)$. Hence ($\mathrm{\bf\tilde R}\equiv\bm\Gamma_0\mathrm{\bf
 R}\bm\Gamma_0^{-1})$,
 \[
 \bm\Upsilon_+^{{\rm (reg)}}(k_1)= \left[\bm\Upsilon_+(k)-\frac{{\rm
 Res}\left[\bm\Upsilon_+(k),k_1\right]}{k-k_1}\right]_{|k=k_1}=
 2k_1\int{\rm d}t\mathrm{\bf E}^{-1}(k_1)\left({\bf
 1}+\frac{1}{2k_1}\mathrm{\bf D}_+\right)\mathrm{\bf \tilde
 R}\bm\Gamma(k_1)\mathrm{\bf E}(k_1)
 \]
 \[
 +\frac{8{\rm i}}{\alpha}k_1^2\left[\bm\sigma_3,\!\int{\rm d}t\,t\mathrm{\bf
 E}^{-1}(k_1)\mathrm{\bf D}_-\mathrm{\bf \tilde R}\bm\Gamma(k_1)\mathrm{\bf
 E}(k_1)\right]\!+2\!\!\int{\rm d}t\mathrm{\bf E}^{-1}(k_1)\mathrm{\bf D}_-
 \mathrm{\bf\tilde R}\left({\bf 1}+\frac{\bar k_1\mathrm{\bf\tilde D}_-}{(k_1-\bar
 k_1)^2}-\frac{\bar k_1\mathrm{\bf \tilde D}_+}{(k_1+\bar k_1)^2}
 \right)\mathrm{\bf E}(k_1).
 \]
 Taking now into account that
 \[
 |\tilde 1\rangle=\exp\left[\int_0^z{\rm
 d}z'\Omega(k_1)\bm\sigma_3\right]|(p_1,p_2)^T\rangle
 \]
 and $p_1/p_2=\exp(\tau_0+{\rm i}\psi_0)$, we find from
 (\ref{3.10})
 \begin{eqnarray*}
 \tau_{0z}&=&\epsilon\int\frac{{\rm d}\tau}{c_-^2}\left\{\left[\frac{{\rm
 i}}{4}\alpha c_+-\frac{2\tau}{w}\left(k_1^3{\rm e}^\tau+\bar
 k_1^3{\rm e}^{-\tau}\right)\right]\rho_--2\tilde t\left(k_1^3{\rm e}^\tau+\bar
 k_1^3{\rm e}^{-\tau}\right)\rho_+\right\},\\
 \psi_{0z}&=&2{\rm i}\epsilon\tilde t\int\frac{{\rm d}\tau}{c_-^2}
 \left(k_1^3{\rm e}^\tau-\bar
 k_1^3{\rm e}^{-\tau}\right)\rho_+-{\rm i}\frac{\epsilon}{w}
 \int\frac{{\rm d}\tau}{c_-^2}\left[\frac{1}{2}\left(k_1^2+\bar
 k_1^2\right)c_++|k_1|^2c_--2\tau\left(k_1^3{\rm e}^\tau-\bar
 k_1^3{\rm e}^{-\tau}\right)\right]\rho_-.
 \end{eqnarray*}
 Finally, differentiation of $\tilde t$ and $\tilde \psi$
 (\ref{2.37}) in $z$ gives equations for the soliton centre
 $\tilde t$ and phase $\tilde\psi$:
 \begin{eqnarray}\label{3.15}
 \tilde t_z&=&v-2\frac{\epsilon}{w}\int\frac{{\rm d}\tau}{c_-^2}
 \left[\frac{{\rm i}}{8}\alpha c_+-\frac{\tau}{w}\left(k_1^3{\rm
 e}^\tau+\bar k_1^3{\rm e}^{-\tau}\right)\right]\rho_-,\\
 \tilde\psi_z&=&\frac{1}{2}\left(v^2+w^2\right)+\frac{\epsilon}{w}
 \int\frac{{\rm
 d}\tau}{c_-^2}\left\{\frac{2\tau}{w}\left[k_1^3(v+{\rm i}w){\rm
 e}^\tau+\bar k_1^3(v-{\rm i}w){\rm e}^{-\tau}\right]-\frac{{\rm
 i}}{4}c_+-{\rm i}|k_1|^2c_-\right\}\rho_-.\label{3.16}
 \end{eqnarray}
 It is remarkable that the even perturbation $(\rho_-=0)$
 influences the soliton velocity and width only, while the odd
 perturbation $(\rho_+=0)$ does not modify these parameters and does modify
 soliton position and phase.

 \section{Subpicosecond DM-soliton in the adiabatic approximation}

 Now we invoke equations (\ref{3.14})--(\ref{3.16}) to analyze the DM-soliton
 parameters in the
 adiabatic approximation with $\epsilon=\mu^2$ and $R=r+{\rm i}\alpha r_t$.
 Substituting the soliton solution $u_s$
 (\ref{2.40}) into (\ref{1.12}), we obtain
 \be\label{4.1}
 r=-16{\rm i}w^7|k_1|^4\frac{f(\tau)}{c_+^3}\,{\rm e}^{{\rm i}\psi},
 \ee
 where
 \be\label{4.2}
 f(\tau)=\frac{{\rm e}^{4\tau}+{\rm
 e}^{-4\tau}+6}{c_+^2c_-^2}+12\left(\frac{4}{c_-^4}-\frac{1}{c_+^4}\right)
 +4\frac{{\rm e}^{2\tau}+{\rm
 e}^{-2\tau}}{c_+c_-}\left(\frac{1}{c_+^2}-\frac{5}{c_-^2}\right)+\frac{1}{|k_1|^2}
 \left(\frac{1}{c_+^2}-\frac{2}{c_-^2}\right).
 \ee
 It follows immediately from (\ref{4.2}) that
 \[
 \rho_+=[(r+{\rm i}\alpha r_t)(\tau)+(\bar r-{\rm i}\alpha\bar
 r_t)(-\tau)]
 \exp(-{\rm i}\psi)=0.
 \]
 Hence, the
 DM-soliton preserves its velocity $v$ and width $w^{-1}$ in the
 adiabatic approximation. Without loss of generality, we pose in
 what follows $v=0$.

 As regards $\rho_-$, it is represented in the form
 \[
 \rho_-=-64{\rm i}w^7|k_1|^4\bar
 k_1^2\left[(1+\beta^2)\frac{f}{c_+^3}-(1-\beta^2)\left(\frac{f}{c_+^3}
 \right)_\tau\right],
 \]
 where $\beta=k_1/\bar k_1=\exp(2{\rm i}\theta)$. In fact, we need
 not calculate the term $(f/c_+^3)_\tau$ because the integration
 in $\tau$ in (\ref{3.15}) permits to transform the integral with
 $f_\tau$, by means of integration by parts, into integrals with
 $f$. As a result, we obtain the following formula for $\tilde
 t_z$:
 \be\label{4.3}
 \tilde t_z=-32\mu^2\alpha w^6|k_1|^4\bar k_1^4\int\frac{{\rm
 d}\tau}{c_+^3c_-^3}\Bigl\{1+4\beta^2+\beta^4+3\beta^2\left(\beta{\rm
 e}^{2\tau}+\beta^{-1}{\rm
 e}^{-2\tau}\right)
 \ee
 \[
 +\frac{4\tau}{1-\beta^2}\left[-1+2\beta^2+2\beta^4
 -\beta^6+\beta^3\left(\beta^2{\rm
 e}^{2\tau}+\beta^{-2}{\rm e}^{-2\tau}\right)\right]\Bigr\}f.
 \]
 Calculation of integrals in (\ref{4.3}) is straightforward and
 gives
 \be\label{4.4}
 \tilde
 t_z=32\mu^2\alpha\left(\frac{2|k_1|^2}{\alpha}\right)^6\left[\frac{1}{12|k_1|^4}
 P_1(\theta)-\frac{16\theta}{\alpha w}P_2(\theta)\right],
 \ee
 \be\label{4.5}
 P_1(\theta)=7\cos(8\theta)+244\cos(4\theta)+278, \qquad
 P_2(\theta)=8\cos(8\theta)+35\cos(4\theta)+44.
 \ee
 The condition $v=0$ implies in accordance with (\ref{2.36}) and
 $k_1=|k_1|\exp({\rm i}\theta)$ the following expressions for
 $|k_1|^2$ and $w$:
 \[
 4|k_1|^2=(1+\alpha^2w^2)^{1/2}, \qquad \alpha w=\tan(2\theta).
 \]
 Therefore, equation (\ref{4.4}) takes the form
 \be\label{4.6}
 \tilde
 t_z=\frac{2\mu^2}{3(\alpha\cos(2\theta))^5}\left(P_1(\theta)\cos(2\theta)-\frac
 {12\theta P_2(\theta)}{\sin(2\theta)}\right).
 \ee
 Now it is reasonable to take up a question of how to minimize a
 perturbation-induced deviation of the soliton centre
 position from that of the unperturbed soliton. Evidently, $\tilde
 t_z$ will be strictly zero in zeros of the curve
 \be\label{4.7}
 Y(\theta)=24\theta P_2(\theta)-P_1(\theta)\sin(4\theta).
 \ee

\begin{figure}
 \begin{center}
 \includegraphics[scale=0.3, angle=270]{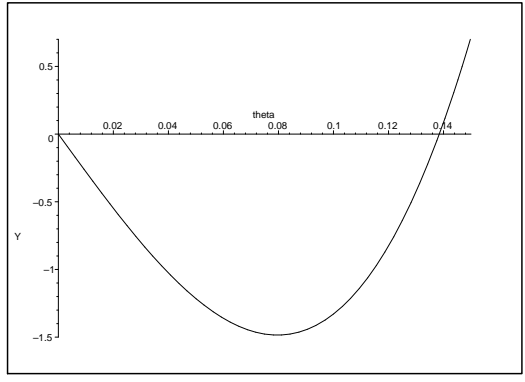}
 \caption{The curve (\ref{4.7})
 intersects the horizontal axis in the point $\theta_1=0.138$.
 The soliton width and the nonlinearity dispersion parameter $\alpha$
 obeying the condition $\alpha w=\tan(2\theta_1)$ provide the
 perturbation-free-like propagation of the DM-soliton.}
 \label{fig.2}
 \end{center}
 \end{figure}

 The plot of the curve (\ref{4.7}) is shown in Figure \ref{fig.2}. It is
 seen that there exists indeed the nontrivial point $\theta_1$ of
 intersection of the curve with the horizontal axis. Numerical
 solution of equation (\ref{4.7}) gives $\theta_1=0.138$. It means
 that if the nonlinearity dispersion $\alpha$ and the inverse
 soliton width $w$ obey the condition $\alpha
 w=\tan(2\theta_1)=0.273$, the perturbation would not
 disturb dynamics of the soliton centre. Note that such a
 compensation is impossible for the NLS DM-soliton.

 Just in the same manner we can calculate a soliton phase
 variation from equation (\ref{3.16}). However, for the single
 soliton this parameter is less informative as compared with the
 soliton centre variation, and we do not exemplify here the
 results.

 \section{Perturbation-induced radiation by DM-soliton}

 For treating radiation effect, we should go beyond the adiabatic
 approximation. It means that we should abandon the condition $\mathrm{\bf
 G}=
 {\bf 1}$ valid in the adiabatic approximation and find the
 function $b(k)$ entering the matrix $\mathrm{\bf G}$
 (\ref{2.18a}). Indeed, the spectral density of the radiation
 energy emitted by the perturbed soliton is given by
 \cite{Lashkin}
 \[
 \mathcal{E}_{{\rm rad}}(k)=-\frac{1}{2\pi\alpha
 k^2}\ln\left[1-{\rm sgn}(k^2)|b(k)|^2\right].
 \]
 The function $b(k)$ accounting for radiation is of the order of
 $\mu^2$. Hence, for $\mu^2\ll 1$ we have
 \[
 \mathcal{E}_{{\rm rad}}(k)=\frac{|b(k)|^2}{2\pi\alpha |k|^2}.
 \]
 Then the spectral density of radiation power is written as
 \be\label{5.1}
 W(k)=\frac{{\rm d}}{{\rm d}z}\mathcal{E}_{{\rm rad}}(k)=\frac{1}
 {\pi\alpha |k|^2}{\rm Re}(b\;\bar b_z).
 \ee

 Now we turn to the evolution equation (\ref{3.8}) for the matrix
 ${\rm\bf G}$ which determines the RH problem (\ref{2.18}). Taking
 the evolution functionals $\bm\Pi_\pm$ in the form
 (\ref{3.9}), we immediately obtain the equation for $\bar b$:
 \be\label{5.2}
 \bar b_z=2\Omega(k)\bar b-\mu^2\Upsilon_{+12}.
 \ee
 Note that $\Upsilon_+=\Upsilon_-$ in the order of $\mu^2$, in
 accordance with equation (\ref{3.6}). The matrix element
 $\Upsilon_{+12}$ is calculated via (\ref{3.12}) where we drop the terms which
 rapidly decrease at infinities:
 \be\label{5.3}
 \Upsilon_{+12}=4\pi\frac{k^3}{|k_1|^2}\frac{(1+\alpha^2w^2)^3}{\alpha^7w}{\rm
 e}^\theta\cos(2\theta)\exp\left[-\frac{{\rm
 i}}{2}(k^2-1)\frac{\tau_0}{w}+{\rm
 i}\phi_0\right]\exp\left(\frac{{\rm
 i}}{2}w^2z\right)H(k,\theta){\rm sech}\frac{\pi p}{2},
 \ee
 where
 \begin{eqnarray*}
 H(k,\theta)&=&gk_1\left\{2p\left[-17-p^2+(p^2-3)\cos(4\theta)\right]-2{\rm
 i}\left[-19-3p^2+(3p^2-1)\cos(4\theta)\right]\right\}\\
 &-&g^{-1}\bar
 k_1\left\{p\left[59-p^2+(p^2-79)\cos(4\theta)\right]+5{\rm
 i}\left[9-3p^2+(3p^2-29)\cos(4\theta)\right]\right\},
 \end{eqnarray*}
 \[
 g=\frac{k^2-\bar k_1^2}{k^2-k_1^2}, \qquad p=\frac{4k^2-1}{\alpha
 w}.
 \]
 Let us remind that $\alpha w=\tan(2\theta)$. Inserting
 (\ref{5.2}) and (\ref{5.3}) into (\ref{5.1}), averaging in $z$
 over the chromatic dispersion map period $\ell$ and dropping fast
 oscillating term, we obtain the following expression (of the order of
 ${\cal O}(\mu^4)$) for the
 spectral density of radiation power:
 \be\label{5.4}
 W(k)=\left(\frac{2}{\alpha}\right)^{10}\frac{\pi\mu^4k^4}{\ell
 w^2}\frac{(1+\alpha^2w^2)^4}{\left[\alpha^2w^2+(4k^2-1)^2\right]^2}|H(k,\theta)|^2
 {\rm e}^{2\theta}{\rm
 sech}^2\left(\frac{\pi}{2}\frac{4k^2-1}{\alpha w}\right).
 \ee

 \begin{figure}
 \begin{center}
 \includegraphics[scale=0.3, angle=270]{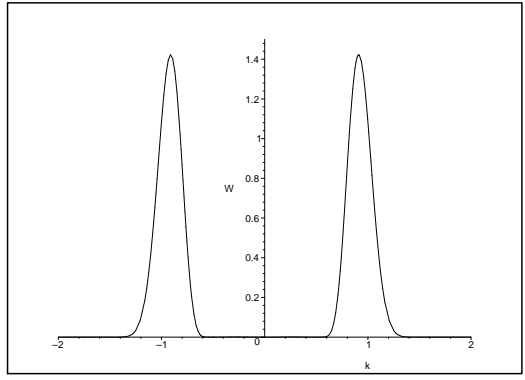}
 \caption{Typical dependence of the spectral density of radiation power
 $W(k)$ (in arbitrary units) on wave number. Two peaks correspond to forward
 and backward radiation emitted by perturbed soliton.}
 \label{fig.3}
 \end{center}
 \end{figure}
 Figure \ref{fig.3} demonstrate a typical dependence of the
 radiation power on the parameter $k$ (which is proportional to the wave
 number). It has two peaks which correspond to forward and
 backward radiation. At the same time, radiation power depends
 only slightly on the nonlinearity dispersion parameter $\alpha$,
 this dependence being detectable for $\alpha\gtrsim 1$.

 \section{Conclusion}

 In this paper, dynamics of a subpicosecond optical pulse in a
 dispersion-managed fibre links has been studied. The analysis has
 been performed in the limit, where the nonlinearity and average
 chromatic dispersion affect the pulse evolution on a much larger
 distance than the local dispersion does. We have derived the
 Gabitov--Turitsyn equation for the subpicosecond DM-soliton
 amplitude. For the case of the weak chromatic dispersion
 management, this equation has been reduced to a perturbed
 modified NLS equation. In the framework of the RH problem
 associated with the MNLS equation, a perturbation theory for the
 MNLS soliton has been elaborated. A possibility has been
 discovered to control the perturbation-induced shift of the
 soliton centre position choosing properly the soliton width and
 nonlinearity dispersion parameter. Spectral density of radiation
 power emitted by the DM-soliton has been found.

 We believe the integrable MNLS equation serves as a true basic
 model to account for the Raman self-frequency shift and
 third-order chromatic dispersion. Both these effects have been
 analysed for the NLS DM-soliton in papers \cite{L-A,L-K}. An
 example of the Raman effect action on the usual MNLS soliton
 was considered in \cite{Af}. As regards the third-order
 dispersion effect for the MNLS soliton, the standard perturbation theory
 cannot capture essential features of the soliton dynamics, and
 a perturbation method that goes beyond all orders is needed
 (see \cite{Chen} for the NLS soliton). This challenging problem is
 outside the scope of the present paper.
 \bigskip

 \noindent{\bf Acknowledgments}\\
 \medskip

 \noindent This work was partly supported by the Belarussian Foundation for
 Fundamental Research under Grant No. $\Phi$04P-071.

 \end{document}